# Orbital and Spin Excitations in Cobalt Oxide

Z. Yamani,[a*], W.J.L. Buyers,[a,b] R.A. Cowley[c] and D. Prabhakaran[c]

[a]*CNBC, National Research Council, Chalk River Laboratories, ON K0J 1J0, Canada*

[b]*Canadian Institute for Advanced Research, Toronto, ON M5G 1Z8, Canada*

[c]*Clarendon Laboratory, University of Oxford, Parks Road, Oxford, OX1 3PU, UK*

**Abstract**

By means of neutron scattering we have determined new branches of magnetic excitations in orbitally active CoO ($T_N$=290 K) up to 15 THz and for temperatures from 6 K to 450 K. Data were taken in the (111) direction in six single-crystal zones. From the dependence on temperature and **Q** we have identified several branches of magnetic excitation. We describe a model for the coupled orbital and spin states of $Co^{2+}$ subject to a crystal field and tetragonal distortion.

The magnetic properties of cobalt compounds have recently attracted attention [1,2] because the Co ion may have transitions between high and low spin states giving rise to charge ordering, magnetic ordering and superconductivity. The recent interest in exotic oxides displaying superconductivity or orbital order emphasizes the importance of understanding simple oxides such as CoO. An earlier study provided an incomplete picture [3].

Cobaltous oxide is a face-centred antiferromagnet in which (111) ferromagnetic sheets of spins stack antiferromagnetically along [111] directions. At any wave-vector transfer, **Q**, up to as many as four spin wave modes may arise from domain structure for any particular spin or orbit transition due to the domain structure. The nearest neighbour exchange is frustrated, contributing no molecular field, and it is the next nearest neighbour exchange that breaks the symmetry below $T_N$ = 290 K. The nature of the order and fluctuations is still controversial [4,5].

We have made high-resolution measurements of the magnetic excitations in the (HHL) plane of a high-quality crystal of CoO at 6 K, 320 K and 450 K. In the first set of experiments, with a focusing PG(002) monochromator and flat PG analyzer with $E_f$=3.52 THz, a resolution of 1.2 THz was achieved at 10 THz energy transfer. In a second set of measurements, with a Be(002) monochromator and PG(002) analyzer set to $E_f$=7.37 THz, a resolution of 0.8 THz was achieved for excitations at 10 THz.

With both configurations we resolved four peaks at 6 K in the excitation spectrum between 4 and 12 THz as shown in Fig. 1 for different magnetic zones. In a recent study only two broad modes in the same spectral range were detected [6]. Analysis including the $Co^{2+}$ magnetic form factor shows that the peak at 9.5 THz is magnetic in origin. The intensities of the peaks at 6.5 THz and 7.6 THz decrease with Q by less than the form factor, suggesting these peaks are partially magnetic. The peak at 4.8 THz has a substantial phonon component and lies close in frequency to the TA zone boundary phonon. This phonon has the same cross-section at (1/2 1/2 3/2), (3/2 3/2 1/2) and (5/2 5/2 1/2) zones centres of Fig. 1, as observed, so that we conclude that the lowest peak carries largely phonon amplitude. The intensities of the three peaks at 6.5, 7.6 and 9.5 THz all have substantial magnetic weight with the lower two showing a mix with vibrational amplitude. In the same freequency region lie the transverse optic and longitudinal acoustic phonons at ~7.6 THz and ~8.3 THz [3].

---

[*] Corresponding author. Tel.: +1-613-584-3311; fax: +1-613-584-4040; e-mail: zahra.yamani@nrc.gc.ca.

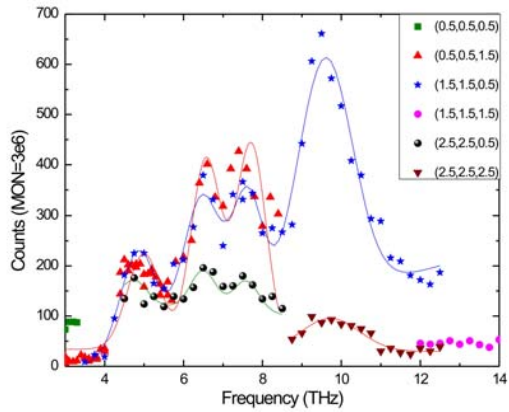

Fig. 1: The magnetic excitations at different zone centres at 6 K fitted to gaussian profiles

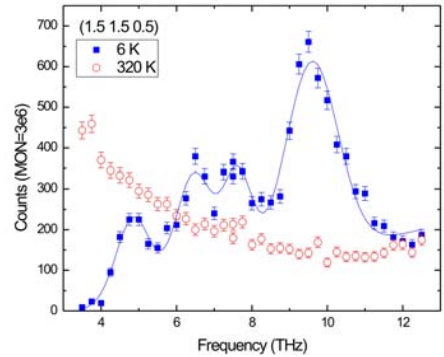

Fig. 2 Sharp low temperature spin waves are replaced above $T_N$ by paramagnetic scattering.

The temperature dependence of the excitations is shown in Fig. 2. A comparison of the data at 320 K (above $T_N$) and 6 K shows that the peak at 9.5 THz disappears on increasing temperature to 320 K. This confirms that this peak is indeed magnetic in origin. The intensities of the other peaks also decrease becoming unobservable relative to the paramagnetic scattering, suggesting they are largely magnetic.

Within each domain there are several bands of spin excitations to states controlled by the exchange, spin-orbit and tetragonally distorted crystal field. Numerical calculations of these modes with a Hund's rule model suggests that the out-of-plane domains have similar frequencies and so the domain structure accounts for three distinct modes. This leads to a different model than that proposed in recent low-resolution experiments [6]. Our analysis shows that even a 'simple' transition metal oxide can exhibit complex phenomena in its spin excitations.